\documentclass[12pt]{article}

\usepackage{amssymb}
\usepackage{amsmath}
\usepackage{slashed}
\usepackage{setspace}
\usepackage{indentfirst}

\usepackage{url}

\begin{document}

\begin{titlepage}
\begin{center}

\vspace*{25mm}

\begin{spacing}{1.7}
{\Large\bf Inflaton as a pseudo-Nambu-Goldstone boson}
\end{spacing}

\vspace*{25mm}

{\large
Noriaki Kitazawa
}
\vspace{10mm}

Department of Physics, Tokyo Metropolitan University,\\
Hachioji, Tokyo 192-0397, Japan\\
e-mail: noriaki.kitazawa@tmu.ac.jp

\vspace*{25mm}

\begin{abstract}
In the realistic model of cosmic inflation
 the inflaton potential should be flat and stable under quantum corrections.
It is natural to imagine that there is some symmetry behind
 and an idea of
 the inflaton as a Nambu-Goldstone boson of spontaneous breaking of some symmetry
 has been examined.
We give a general formulation of this idea
 using the non-linear realization of Nambu-Goldstone boson in low-energy effective theory
 with some explicit symmetry breaking to generate non-trivial potential.
The potential is naturally a simple function, typically the mass term of inflaton,
 and the scenario of ``warm inflation'' should necessarily be applied
 under the present observational constraints.
We investigate
 the generation mechanism of necessary thermal dissipation term in inflaton field equation
 for ``warm inflation'' without large thermal corrections to inflaton potential.
A simple numerical analysis is given to investigate the viability of this scenario.
\end{abstract}

\end{center}
\end{titlepage}


\section{Introduction}
\label{sec:introduction}

The slow-roll inflation
 by the potential energy by almost flat potential as a function of inflaton field
 is considered to be most probable.
In many candidates of the models of inflation potential
 (see \cite{Martin:2013tda}, for example, for the status of various models)
 the common problem is the stability under quantum corrections
 by the fields which necessarily interact with inflaton for the reheating.
In general
 the potential of scalar field with some interactions get large quantum corrections,
 which courses the naturalness problem in many places in particle physics and cosmology.
The ideas of
 ``natural inflation'' \cite{Freese:1990rb,Adams:1992bn}
 and its improvements \cite{Croon:2014dma,Croon:2015fza},
 ``warm little inflaton'' \cite{Bastero-Gil:2016qru,Benetti:2016jhf} and
 ``minimal warm inflation''
 \cite{Berghaus:2019whh,Laine:2021ego,Visinelli:2011jy,Kamali:2019ppi}
 are proposed as a solution of this problem
 by assuming the inflaton as a pseudo-Nambu-Goldstone boson.
Since the exact Nambu-Goldstone boson can not have potential for symmetry reason,
 the potential of pseudo-Nambu-Goldstone boson
 due to certain explicit soft symmetry breaking effect
 is not suffered by large quantum corrections.
 
On the other hand, because of the symmetry reason,
 the form of such a potential is restricted to be a simple one,
 which is not acceptable in normal inflation scenario
 by too large tensor-to-scalar ratio as a necessary result of large field inflation.
In the scenario of warm inflation
 \cite{Berera:1995ie,Berera:1995wh}
 with radiation in almost thermal equilibrium
 even under the accelerated expansion of the universe,
 such a simple inflaton potential is acceptable.
Since the additional friction by
 thermal dissipation term in the inflaton field equation
 due to the interaction with radiation
 keeps the value of the inflaton field small
 as well as realizing enough number of e-folds in slow-roll inflation.

In this letter we give a general arguments of warm inflation
 with an inflaton as pseudo-Nambu-Goldstone boson.
In the next section
 we introduce a general formulation of Nambu-Goldstone field
 in low-energy effective theory with non-linear realization 
 \cite{Weinberg:1978kz,Gasser:1983yg}.
In section \ref{sec:pot-and-diss}
 we discuss some explicit soft symmetry breaking
 to produce inflaton potential and thermal dissipation term in inflation field equation.
In section \ref{sec:numerical}
 we solve the fundamental equations of warm inflation in slow-roll limit
 and give a simple numerical analysis to investigate the viability of this scenario.
In the last section we conclude.

\section{Effective theory with non-linear realization}
\label{sec:effective-theory}

The dynamics of Nambu-Goldstone boson
 at energies much lower than the energy scale of
 corresponding spontaneous symmetry breaking
 is well described by low-energy effective theory
 with non-linear realization
 (see \cite{Coleman:1969sm,Callan:1969sn,Bando:1987br} for general formalism).
In this letter
 we consider only a very simple model of the symmetry breaking
 $\rm{U}(1) \rightarrow {\bf Z}_2$ at the energy scale $f$
 as a minimal case.
The non-linear Nambu-Goldstone field is
\begin{equation}
 U = e^{i \phi/f},
\end{equation}
 where $\phi$ is the Nambu-Goldstone field which transforms as
 $\phi \rightarrow \phi + \theta f$ with $0 < \theta \le 2\pi$ under the U$(1)$.
We assume that
 the spontaneous symmetry breaking is effectively triggered by
 a vacuum expectation value of a scalar field with U$(1)$ charge $2$,
 and the low-energy effective theory is invariant under ${\bf Z}_2$ transformation
 $U \rightarrow -U$ corresponding to $\theta=\pi$.
The low-energy effective Lagrangian of the field is constructed
 in the notion of a derivative expansions
 (we neglect non-trivial background space-time metric for a while).
\begin{equation}
 {\cal L}_{\rm eff}
  = \frac{1}{2} f^2 \partial^\mu U^\dag \partial_\mu U + {\cal O}(\partial^4).
\label{eq:lagrangian-only-phi}
\end{equation}
In this simple model the first term gives kinetic term only
 and the interaction with derivative couplings are included in higher order terms.
There is no potential as a function of $\phi$ without derivatives.
In this letter we investigate low-energy effective theory at the lowest order.

Inclusion of fermions in an U$(1)$ symmetric way
 reveals the non-trivial nature of Nambu-Goldstone boson.
Introduce a left-handed fermion $\psi_L$ with U$(1)$ charge $1$
 and a right-handed fermion $\psi_R$ with U$(1)$ charge $0$,
 then the simplest effective Lagrangian is
\begin{equation}
 {\cal L}_{\rm eff}
  = \frac{1}{2} f^2 \partial^\mu U^\dag \partial_\mu U
  + {\bar \psi} i \slashed{\partial} \psi
  - \mu U {\bar \psi}_L \psi_R - \mu U^\dag {\bar \psi}_R \psi_L,
\label{eq:lagrangian-with-fermion}
\end{equation}
 where $\mu$ is a real parameter with dimension one \cite{Bastero-Gil:2016qru}.
The mass of the fermion becomes clear
 once we redefine the field as $U^\dag \psi_L \rightarrow \psi_L$
 (analogous to ``constituent quark field'').
The effective Lagrangian becomes in this new field
\begin{equation}
 {\cal L}_{\rm eff}
  = \frac{1}{2} f^2 \partial^\mu U^\dag \partial_\mu U
  + {\bar \psi} i \slashed{\partial} \psi
  - \mu {\bar \psi} \psi - \frac{1}{f} \partial_\mu \phi {\bar \psi}_L \gamma^\mu \psi_L.
\label{eq:lagrangian-constituent}
\end{equation}
We see that
 the derivative coupling of the Nambu-Goldstone field $\phi$ with fermion U$(1)$ current,
 and fermion one-loop quantum corrections do not give potential terms of $\phi$
 without derivatives.
The field equation of $\phi$ contains the divergence of fermion U$(1)$ current,
 which can be rewritten as a pseudo-scalar density by using fermion field equation, and
\begin{equation}
 \Box \phi + \frac{\mu}{f} {\bar \psi} i \gamma_5 \psi = 0.
\end{equation} 
Even if we consider a thermal bath of the fermion,
 no effect (thermal mass, dissipation term, for example)
 emerges in the equation of $\phi$,
 because the thermal average of the parity-odd pseudo-scalar density vanishes.
Here, and in this letter,
 we apply the Hartree approximation to evaluate thermal effects
 for a simple and intuitive understanding \cite{Yokoyama:1998ju}.

In this way
 there are no quantum corrections to the potential and no thermal effects
 for Nambu-Goldstone field.
We need to introduce some explicit soft breaking of U$(1)$ symmetry
 whose energy scale is much less than
 the energy scale of spontaneous symmetry breaking $f$.

\section{Inflaton potential and dissipation term}
\label{sec:pot-and-diss}

Consider a spurious scalar field $\Sigma$ with U$(1)$ charge $-2$.
Then, the following potential term is arrowed.
\begin{equation}
 V_{\rm eff} = - \frac{1}{8} B \Sigma U U + {\rm h.c.},
\end{equation}
 where $B \simeq f^3$
 is a real quantity which are determined by the dynamics of
 spontaneous U$(1)$ symmetry breaking.
This form is analogous to
 the explicit chiral symmetry breaking term by current quark masses
 in the low-energy effective Lagrangian for pions
 which are Nambu-Goldstone bosons under the spontaneous chiral symmetry breaking
 by QCD dynamics.
The real vacuum expectation value of the spurious field,
 $\langle \Sigma \rangle = v_\Sigma \ll f$,
 gives the following ${\bf Z}_2$ symmetric potential
\begin{equation}
 V_{\rm eff} = - \frac{1}{4} M^4 \cos(2\phi/f)
 =  - \frac{1}{4} M^4 + \frac{1}{2} \frac{M^4}{f^2} \phi^2
    - \frac{1}{6} \frac{M^4}{f^4} \phi^4
    + {\cal O}((\phi/f)^6),
\label{eq:potential}
\end{equation}
 where $M^4 \equiv B v_\Sigma \simeq f^3 v_\Sigma$.
Since this potential should be considered in the range $\phi \ll f$
 as that in the low-energy effective theory,
 the dominant term is the mass term of $\phi$
 with the mass $m_V^2 \simeq f v_\Sigma$. 
Here, we simply neglect the constant term,
 which belongs to the cosmological constant (or dark energy) problem
 beyond the scape of this work.

The quartic coupling in eq.(\ref{eq:potential})
 gives quantum corrections to the potential,
 but all the correction is higher order in the expansion of $M^4/f^4 \simeq v_\Sigma/f$
 \cite{Weinberg:1978kz,Gasser:1983yg}
 (we need to use dimensional regularization not to break U$(1)$ symmetry).
This expansion is equivalent to
 the derivative expansion in eq.(\ref{eq:lagrangian-only-phi})
 with $\partial^2/f^2 \sim M^4/f^4$
 so that the mass term is the same order of the kinetic term.
Therefore, the inflaton potential under $\phi \ll f$
\begin{equation}
 V \simeq \frac{1}{2} m_V^2 \phi^2
\label{eq:inflaton-potential}
\end{equation}
 is stable under the quantum corrections as long as we stay in the lowest order
 of the expansion in $M^4/f^4 \simeq v_\Sigma/f$.

The generation of dissipation term in the field equation of inflaton for warm inflation
 is more complicated, because the thermal mass, which is the dominant thermal effect,
 must be canceled out.
At least two fermions are required with some tuned parameters.

First consider the thermal correction
 from a single fermion field with the symmetry breaking term
\begin{equation}
 {\cal L}_{\rm eff}^{\rm br}
  = - \varphi {\bar \psi}_L \psi_R + {\rm h.c.}
\end{equation}
 in the effective Lagrangian of eq.(\ref{eq:lagrangian-with-fermion}),
 where $\varphi$ is a spurious field with U$(1)$ charge $1$.
Redefine the field as $U^\dag \psi_L \rightarrow \psi_L$, 
 and introduce a real vacuum expectation value of the spurious field $\varphi$
 as $m \ll f$ and $m \ll \mu$,
 then we have
\begin{equation}
 {\cal L}_{\rm eff}^{\rm br}
 = - m \cos(\phi/f) {\bar \psi} \psi + m \sin(\phi/f) {\bar \psi} i \gamma_5 \psi,
\label{eq:breaking-single-fermion}
\end{equation}
 which should be added to eq.(\ref{eq:lagrangian-constituent}).
The fermion one-loop quantum correction by these interactions
 is independent from the inflaton field
 due to the nature of trigonometric functions
 or the unitarity of non-linear Nambu-Goldstone field,
 and it does not give corrections to the inflaton potential.
The effective mass of the fermion under a background field $\phi$ is
\begin{equation}
 m_{\rm eff} = \mu + m \cos(\phi/f),
\end{equation}
 where we treat the pseudo-scalar density in eq.(\ref{eq:breaking-single-fermion})
 as an interaction term.
The field equation of the inflaton field becomes
\begin{equation}
 \Box \phi + V' + \frac{\mu}{f} {\bar \psi} i \gamma_5 \psi
  - \frac{m}{f} \sin(\phi/f) {\bar \psi} \psi
  - \frac{m}{f} \cos(\phi/f) {\bar \psi} i \gamma_5\psi = 0.
\end{equation} 
The Hartree approximation
 assuming near thermal distribution of fermions with temperature $T \gg \mu$ gives
\begin{equation}
 \langle {\bar \psi} \psi \rangle
 \simeq \frac{1}{6} m_{\rm eff} T^2
 + 4 \mu^2 \frac{d m_{\rm eff}}{d \phi} \frac{C}{\Gamma_{\rm th}} {\dot \phi}
 + {\cal O}((m/f)^2),
\qquad
 \langle {\bar \psi} i \gamma_5 \psi \rangle = 0,
\end{equation}
 where $C$ is a constant of order $10$
 and $\Gamma_{\rm th}^{-1}$ is the time to recover thermal equilibrium
 from the disturbance by small change of the value of $\phi$ in fermion effective mass.
The effective inflaton field equation becomes
\begin{equation}
 \Box \phi + V'
  - \frac{m}{f} \sin(\phi/f)
    \left[ \frac{1}{6} (\mu + m\cos(\phi/f)) T^2
     - 4 \mu^2 \frac{m}{f} \sin(\phi/f) \frac{C}{\Gamma_{\rm th}} {\dot \phi}
    \right]
  \simeq 0.
\label{eq:field-eq-thermal}
\end{equation}
The first term in the square bracket is the thermal mass term
 and the second term is the dissipation term.
We see that the thermal mass is dominant and the dissipation is subleading effect
 as it has been pointed out in \cite{Yokoyama:1998ju}.

Consider two fermions of the same U$(1)$ charges
 and interactions of eq.(\ref{eq:lagrangian-with-fermion}),
 which are connected with the ${\bf Z}_2$ symmetry
 (a combination of ${\bf Z}_2$ from U$(1)$ breaking
  and ${\bf Z}_2$ of the exchange of $\psi_1$ and $\psi_2$) as
\begin{equation}
 \left(
  \begin{array}{c}
   \psi_{1L} \\ \psi_{2L}
  \end{array}
 \right)
 \rightarrow
 e^{i\pi}
 \left(
  \begin{array}{cc}
   0 & i \\
   -i & 0
  \end{array}
 \right)
 \left(
  \begin{array}{c}
   \psi_{1L} \\ \psi_{2L}
  \end{array}
 \right),
\qquad
 \left(
  \begin{array}{c}
   \psi_{1R} \\ \psi_{2R}
  \end{array}
 \right)
 \rightarrow
 \left(
  \begin{array}{cc}
   0 & i \\
   -i & 0
  \end{array}
 \right)
 \left(
  \begin{array}{c}
   \psi_{1R} \\ \psi_{2R}
  \end{array}
 \right)
\end{equation}
 with $U \rightarrow -U$.
The only difference of these two fermions is the sign of the breaking term
\begin{equation}
 {\cal L}_{\rm eff}^{\rm br}
 = - \varphi {\bar \psi}_{1L} \psi_{1R}
   + \varphi {\bar \psi}_{2L} \psi_{2R} + {\rm h.c.},
\end{equation}
 which becomes
\begin{eqnarray}
 {\cal L}_{\rm eff}^{\rm br}
 &=& - m \cos(\phi/f) {\bar \psi}_1 \psi_1
     + m \sin(\phi/f) {\bar \psi}_1 i \gamma_5 \psi_1
\nonumber\\
 &&  + m \cos(\phi/f) {\bar \psi}_2 \psi_2
     - m \sin(\phi/f) {\bar \psi}_2 i \gamma_5 \psi_2,
\label{eq:breaking-double-fermions}
\end{eqnarray}
 in redefined left-handed fermion fields with $\langle \varphi \rangle = m$.
The largest contribution in the thermal mass in eq.(\ref{eq:field-eq-thermal}),
 which is proportional to $m$, is canceled:
\begin{equation}
 \Box \phi + V'
  - \frac{m}{f} \sin(\phi/f)
    \left[ \frac{1}{3} m\cos(\phi/f) T^2
     - 8 \mu^2 \frac{m}{f} \sin(\phi/f) \frac{C}{\Gamma_{\rm th}} {\dot \phi}
    \right]
  \simeq 0.
\label{eq:field-eq-thermal-two-fermions}
\end{equation}
The contributions which are proportional to the even power of $m$ are not canceled,
 and the subleading thermal mass term remains.
We assume for simplicity that $\Gamma_{\rm th}$ is the same for each fermion.

We need more fermions to cancel the remaining thermal mass term.
Introduce the mirror fields ${\tilde \psi}_1$, ${\tilde \psi}_2$ and ${\tilde \varphi}$
 which are copies of the corresponding fields $\psi_1$, $\psi_2$ and $\varphi$,
 respectively, with a new mirror ${\bf Z}_2$ symmetry
 $\psi \leftrightarrow {\tilde \psi}$ and $\varphi \leftrightarrow {\tilde \varphi}$.
We assume that the physics behind the explicit breaking
 results the vacuum expectation values of the spurious fields as
 $\langle \varphi \rangle = m$ and $\langle {\tilde \varphi} \rangle = im$
 (the mirror ${\bf Z}_2$ symmetry is also broken).
Then the additional breaking terms in redefined left-handed fermion fields are
\begin{eqnarray}
 {\cal L}_{\rm eff}^{{\widetilde {\rm br}}}
 &=& - m \sin(\phi/f) {\bar {\tilde \psi}}_1 {\tilde \psi}_1
     - m \cos(\phi/f) {\bar {\tilde \psi}}_1 i \gamma_5 {\tilde \psi}_1
\nonumber\\
 &&  + m \sin(\phi/f) {\bar {\tilde \psi}}_2 {\tilde \psi}_2
     + m \cos(\phi/f) {\bar {\tilde \psi}}_2 i \gamma_5 {\tilde \psi}_2.
\end{eqnarray}
 which should be compared with eq.(\ref{eq:breaking-double-fermions})
 showing different appearance of trigonometric functions.
The cancelation of the thermal mass term in inflaton field equation happens as
\begin{equation}
 \Box \phi + V'
  + 8 \mu^2 \left( \frac{m}{f} \right)^2 \frac{C}{\Gamma_{\rm th}} {\dot \phi}
  \simeq 0,
\end{equation}
 where we assume again that $\Gamma_{\rm th}$ is the same for each fermion.
This mechanism
 utilizing combinations of trigonometric functions
 has been proposed in \cite{Bastero-Gil:2016qru}.
In this way, even if it is possible,
 we need some complicated mechanism to produce dissipation term
 without large thermal effect to the potential.
Another mechanism is proposed in \cite{Berghaus:2019whh}
 where the dissipation is not thermal but topological
 assuming that the inflaton is an axion with topological coupling
 to some Yang-Mills gauge field.
See \cite{Dymnikova:2000gnk,Dymnikova:2001jy} for further possibilities.

We estimate $\Gamma_{\rm th} \simeq h^2 \mu^2/T$ following \cite{Bastero-Gil:2016qru}
 by considering the decay of the fermions in almost thermal equilibrium
 through the interaction
\begin{equation}
 {\cal L}_{\rm int}
  = - h {\bar \psi}_R \Phi^\dag \ell_L + {\rm h.c.},
\end{equation}
 where $h$ is a coupling constant of this Yukawa coupling
 in which $\ell_L$ and $\Phi$
 (they are also in thermal equilibrium) are possibly identified
 as left-handed lepton doublet and Higgs doublet
 in the standard model of particle physics, respectively as in \cite{Levy:2020zfo}.
Then the coefficient of the dissipation term
 is $\Gamma \simeq (C/h^2) (m/f)^2 T$,
 which does not depend on the inflaton field,
 where newly defined $C$ is a constant of the order of $10$.

\section{Dynamics and numerical evaluation}
\label{sec:numerical}

The fundamental equations of warm inflation are the following \cite{Berera:1995wh}.
\begin{equation}
 {\ddot \phi} + (3H + \Gamma) {\dot \phi} + V' = 0,
\end{equation}
\begin{equation}
 3H^2 = \frac{1}{M_p^2} \left( \frac{1}{2} {\dot \phi}^2 + V + \rho_r \right),
\end{equation}
\begin{equation}
 {\dot \rho}_r + 4H\rho_r = \Gamma {\dot \phi}^2.
\end{equation}
Here, $H$ is the Hubble parameter,
 $\rho_r = C_R T^4$ is the energy density of radiation in almost thermal equilibrium
 with $C_R = g_* \pi^2/30$, where the effective degrees of freedom $g_*$
 which includes the fermion fields in previous section,
 and $M_p$ is the reduced Planck mass.
The model in this letter, the inflaton as a pseudo-Nambu-Goldstone boson,
 we have obtained inflaton potential of eq.(\ref{eq:inflaton-potential}),
 $V \simeq m_V^2 \phi^2$/2,
 and dissipation coefficient as $\Gamma \simeq C_T T$
 with $C_T \equiv (C/h^2) (m/f)^2$.
There are conditions
 which should be satisfied among dimension-full quantities
 so that the low-energy effective theory is applicable.
The value of the inflaton field
 should be less than the energy scale of spontaneous U$(1)$ symmetry breaking:
 $\phi \ll f \lesssim M_p$.
Since the mass of inflaton $m_V$
 is generated by the effect of explicit U$(1)$ symmetry breaking,
 it should be less than $f$: $m_V \ll f$ (more precisely $v_\Sigma \ll f$).
The mass of fermions $\mu$ should be less than $f$
 and the energy scale of explicit symmetry breaking $m$
 should be less than $\mu$ and $f$:
 $m \ll \mu \ll f$.
The temperature should be higher than the fermion mass,
 but lower than the energy scale of spontaneous symmetry breaking:
 $\mu \ll T \ll f$.

The slow-roll approximation of the fundamental equations are
\begin{equation}
 {\dot \phi} \simeq \frac{V'}{3H(1 + Q)},
\qquad
 3H^2 \simeq \frac{V}{M_p^2},
\qquad
 4\rho_r \simeq 3 Q {\dot \phi}^2,
\end{equation}
 where $Q \equiv \Gamma/3H$.
These equations are satisfied when the slow-roll conditions
 $\epsilon \ll Q$ and $\eta \ll Q$ are satisfied, where
\begin{equation}
 \epsilon \equiv \frac{M_p^2}{2} \left( \frac{V'}{V} \right)^2
          \simeq 2 \left( \frac{M_p}{\phi} \right)^2,
\qquad
 \eta \equiv M_p^2 \frac{V''}{V} \simeq \epsilon
\label{slow-roll-param}
\end{equation}
 under the present simple potential of eq.(\ref{eq:inflaton-potential}).
In the usual inflation without radiation,
 the inflaton field value must be larger than the Planck scale for slow-roll
 in this kind of simple potential,
 and this large field inflation gives too large value of tensor-to-scalar ratio.
In warm inflation
 the inflaton field value can be smaller than Planck scale
 for slow-roll as far as $Q \gg 1$.
We first analytically solve the slow-roll equations
 and give the formulae for typical values of $\phi$, $H$, $Q$ and $T/H$
 during slow-roll inflation.

The derivative in time $t$
 can be replaced by the derivative in e-folds $N$ with $dt = H^{-1} dN$,
 and we obtain the difference of the values of field at $N=0$ and $N=N_e$
 from slow-roll equations as
\begin{equation}
 \left(\phi(N=0)\right)^{6/5} - \left(\phi(N=N_e)\right)^{6/5}
 \simeq \frac{6}{5} N_e
        \left(
         \frac{144C_R}{C_T^4} M_p^4 m_V^2
        \right)^{1/5}.
\end{equation}
This gives
 typical values of inflaton field and Hubble parameter during slow-roll inflation as
\begin{equation}
 \phi_{\rm typ} \equiv 
  \left(
   \frac{6}{5} N_e
  \right)^{5/6}
  \left(
   \frac{144C_R}{C_T^4} M_p^4 m_V^2
  \right)^{1/6},
\qquad
 H_{\rm typ} \simeq \frac{1}{\sqrt{6}} \frac{m_V}{M_p} \phi_{\rm typ}.
\end{equation}
The typical value of $Q$ is given in a complicated form as
\begin{equation}
 Q_{\rm typ}
 \simeq
   \sqrt{\frac{2}{3}}
   \left(
    \frac{144^2}{(4\sqrt{6})^5} \frac{C_T^{12}}{C_R^3}
    \left( \frac{M_p}{f} \right)^{18}
    \left( \frac{f}{m_V} \right)^6
    \left( \frac{f}{\phi_{\rm typ}} \right)^{12}
   \right)^{1/15},
\end{equation}
 and this gives the typical value of $T/H$ as
\begin{equation}
 \frac{T_{\rm typ}}{H_{\rm typ}} = \frac{3}{C_T} Q_{\rm typ}.
\end{equation}
Roughly speaking both $\phi_{\rm typ} \ll f$ and $Q_{\rm typ} \gg 1$ requires
\begin{equation}
 \frac{m_V}{M_p} \ll C_T^2
\label{eq:small-mV}
\end{equation}
 namely a flat potential.

If the value of $Q_{\rm typ}$ is large enough,
 the typical value of $T/H$ is large
 and the contribution of thermal fluctuations to the scalar (or curvature) perturbations
 dominates over the contribution of quantum fluctuations
 as it can be seen in the formula of the power spectrum
 \cite{Hall:2003zp,Ramos:2013nsa,Bastero-Gil:2014jsa,Visinelli:2014qla}
\begin{equation}
 P_{\cal R} \simeq
  \left( \frac{Q}{2\sqrt{2}\pi} \frac{H}{M_p} \right)^2 \frac{1}{\epsilon}
  \left[
   1 + \sqrt{3 \pi Q} \frac{T}{H}
  \right]
  G(Q),
\end{equation}
 where we assume $Q \gg 1$ and no inflaton particles in thermal equilibrium and
\begin{equation}
 G(Q) = 1 + 0.335Q^{1.364} + 0.0185 Q^{2.315}
\end{equation}
 which is given in \cite{Bastero-Gil:2016qru,Benetti:2016jhf}
 for the case of $\Gamma \propto T$.
All the quantities in this formula of power spectrum
 are evaluated at the time of $k/a(t)H(t)=1$,
 and it is a function of wave number $k$ in this sense.
The spectral index of this power spectrum is approximately given by
\begin{equation}
 n_s = 1 - \frac{6}{5} \frac{\epsilon}{Q}
       + \frac{d \ln G}{d \ln Q} \cdot \frac{4}{5} \frac{\epsilon}{Q}
\label{spectral-index}
\end{equation}
 in the present particular form of the inflaton potential and dissipation term.

A method as a first attempt to evaluate
 whether this scenario can be realistic or not
 is to choose a set of concrete numerical values of parameters
 and calculate the values of important quantities.
For simplicity
 we assume that U$(1)$ symmetry is spontaneously broken at Planck scale,
 namely $f \simeq M_p \simeq 2.4 \times 10^{18}$ GeV,
 which could be understood as accepting a view point
 that the quantum gravity should not allow a global symmetry.
We choose some natural values of dimensionless parameters as
 $N_e = 60$, $g_* = 100$, $C = 10$ and $h = 2.5 \times 10^{-9}$.
This very small coupling is required
 for enough light fermions to be in thermal equilibrium
 (see the definition of $C_T$ under $f \simeq M_p$).
A possible problem is the thermalization
 with slow-rate production and decay of fermions.
The value of $m_V$ is chosen to be small enough as
\begin{equation}
 \frac{m_V}{M_p} = C_T^2 \times 10^{-8}
\end{equation}
 according to eq.(\ref{eq:small-mV}),
 which makes $\phi_{\rm typ}$ and $Q_{\rm typ}$ independent from $m$.
We obtain
\begin{equation}
 \frac{\phi_{\rm typ}}{f} \simeq 0.3,
\qquad
 Q_{\rm typ} \simeq 10^3,
\end{equation}
 which indicate that
 the low-energy effective theory is applicable
 and the slow-roll condition is satisfied.
The resultant value $\epsilon_{\rm typ} \simeq 20$ is large,
 but $\epsilon_{\rm typ}/Q_{\rm typ} \simeq 0.01$ is small.
We choose $m = M_p \times 10^{-10} \simeq 2 \times 10^8$ GeV
 so that the typical value of Hubble parameter during inflation
 is small enough, $H < 10^9$ GeV,
 which is the condition for stability of the electroweak Higgs potential
 in the standard model of particle physics \cite{Kohri:2016wof}.
Then we obtain
\begin{equation}
 H_{\rm typ} \simeq 8 \times 10^5 \, {\rm GeV},
\qquad
 T_{\rm typ} \simeq 2 \times 10^{11} \, {\rm GeV},
\end{equation}
 which give large $T/H \simeq 3 \times 10^5$
 and small $T/f \simeq 9 \times 10^{-8}$ as they should be.
The fermion mass $\mu$,
 which does not explicitly appear in physical quantities,
 should take some value satisfying $m < \mu <T_{\rm typ}$
 for thermal equilibrium fermions.
The mass of inflaton is $m_V \simeq 6 \times 10^6$ GeV,
 which corresponds to $\langle \Sigma \rangle = v_\Sigma \simeq 0.02$ MeV.

These theoretically consistent values of parameters,
 which properly realize the hierarchy of scales, give
 the approximate values of the amplitude and spectral index of scalar perturbations as
 $A_s \simeq 2 \times 10^{-9}$ and $n_s \simeq 1$,
 which are acceptable in comparison with those obtained by observations,
 $A_S = (2.10 \pm 0.03) \times 10^{-9}$ and $n_s = 0.965 \pm 0.004$
 by \cite{Planck:2018vyg}, for example, within this simple analysis for order estimates.
We observe, however, that
 it could be difficult to obtain the value $n_s < 1$ in this scenario.
In low-energy effective theory with $\phi \ll f \lesssim M_p$,
 the formulae of slow-roll parameters in eq.(\ref{slow-roll-param})
 indicate large slow-roll parameters,
 and a large value of $Q$ is necessary for slow-roll inflation
 ($\epsilon / Q \ll 1$ and $\eta / Q \ll 1$) with enough number of e-folds.
On the other hand,
 in the approximate formula of spectral index, eq.(\ref{spectral-index}),
 the third term dominates over the second term for $Q \gtrsim 10$
 in the right-hand side, and $n_s$ can not be less than one,
 which accords with the analyses of chaotic warm inflation in
 \cite{Bastero-Gil:2016qru,Bastero-Gil:2014jsa}.
Since we can not exclude this scenario by this simple analysis of this work,
 further precise and detailed analyses are necessary.
It could be possible that this scenario,
 a general scenario of inflaton as a pseudo-Nambu-Goldstone boson 
 in low-energy effective theory,
 would be hard to reconcile with observation in the end.

\section{Conclusions}
\label{sec:conclusions}

We have examined the possibility that the inflaton is a pseudo-Nambu-Goldstone boson.
Starting with
 an introduction of low-energy effective theory of Nambu-Goldstone boson
 in non-linear realization
 and showing no quantum corrections to the potential and no thermal effects,
 the way to generate a non-trivial potential and thermal effects
 through the explicit breaking of the symmetry has been reviewed.
Though we could choose more general pattern of spontaneous breaking
 of the symmetry group $G$ to the group $H$ with Nambu-Goldstone fields
 living in the space $G/H$,
 as it has been investigated in \cite{Croon:2015fza},
 we chose U$(1)$ symmetry breaking as the simplest case,
 which is enough to understand essential facts.
The result of this choice
 gives a simple model which is similar to the model in \cite{Bastero-Gil:2016qru},
 but the form of fermion interactions and inflaton potential is more restricted
 for symmetry reason.
We have solved the slow-roll equations
 taking care of the hierarchy of scales in this model,
 and shown that reasonable value of $A_s$ and $n_s$ in the power spectrum
 of scalar perturbations can be obtained with some natural values of parameters
 within our simple analysis for order estimates.
A simple observation, however, would indicate that
 the value $n_s < 1$ could be difficult to be realized
 because of large value of $Q$ is required in low-energy effective theory
 for slow-roll inflation with enough number of e-folds.
Further precise and detailed analyses are necessary
 to clarify the viability of a general scenario of
 inflaton as a pseudo-Nambu-Goldstone boson in low energies effective theory.

In the model building point of view,
 the dynamics of spontaneous U$(1)$ symmetry breaking
 and the origin of explicit breaking (vacuum expectation values of spurious fields)
 should be investigated.
In the phenomenological point of view
 the effect of the derivative coupling of inflaton to fermion U$(1)$ current
 (the last term in eq.(\ref{eq:lagrangian-constituent})),
 which pretends a chemical potential term under the time dependence of inflaton,
 should be investigated.
We need to use thermal field theory beyond Hartree approximation
 for this investigation as well as to obtain more precise form of the dissipation term
 and also to clarify the role of parity-odd pseudo-scalar densities.

An interesting possibility that
 this inflaton becomes cold dark matter,
 which has been proposed in \cite{Levy:2020zfo},
 is worth to be examined within the restriction by symmetry.
We need to investigate the end of inflation
 and reheating process probably by numerical simulations.

\section*{Acknowledgments}

This work was supported in part by JSPS KAKENHI Grant Number 19K03851.


\begin{thebibliography}{99}


\bibitem{Martin:2013tda}
J.~Martin, C.~Ringeval and V.~Vennin,
``Encyclop\ae{}dia Inflationaris,''
Phys. Dark Univ. \textbf{5-6} (2014), 75-235
[arXiv:1303.3787 [astro-ph.CO]].

\bibitem{Freese:1990rb}
K.~Freese, J.~A.~Frieman and A.~V.~Olinto,
``Natural inflation with pseudo - Nambu-Goldstone bosons,''
Phys. Rev. Lett. \textbf{65} (1990), 3233-3236.
\bibitem{Adams:1992bn}
F.~C.~Adams, J.~R.~Bond, K.~Freese, J.~A.~Frieman and A.~V.~Olinto,
``Natural inflation: Particle physics models,
 power law spectra for large scale structure, and constraints from COBE,''
Phys. Rev. D \textbf{47} (1993), 426-455
[arXiv:hep-ph/9207245 [hep-ph]].
\bibitem{Croon:2014dma}
D.~Croon and V.~Sanz,
``Saving Natural Inflation,''
JCAP \textbf{02} (2015), 008
[arXiv:1411.7809 [hep-ph]].
\bibitem{Croon:2015fza}
D.~Croon, V.~Sanz and J.~Setford,
``Goldstone Inflation,''
JHEP \textbf{10} (2015), 020
[arXiv:1503.08097 [hep-ph]].

\bibitem{Bastero-Gil:2016qru}
M.~Bastero-Gil, A.~Berera, R.~O.~Ramos and J.~G.~Rosa,
``Warm Little Inflaton,''
Phys. Rev. Lett. \textbf{117} (2016) no.15, 151301
[arXiv:1604.08838 [hep-ph]].
\bibitem{Benetti:2016jhf}
M.~Benetti and R.~O.~Ramos,
``Warm inflation dissipative effects: predictions and constraints from the Planck data,''
Phys. Rev. D \textbf{95} (2017) no.2, 023517
[arXiv:1610.08758 [astro-ph.CO]].

\bibitem{Berghaus:2019whh}
K.~V.~Berghaus, P.~W.~Graham and D.~E.~Kaplan,
``Minimal Warm Inflation,''
JCAP \textbf{03} (2020), 034
[arXiv:1910.07525 [hep-ph]].
\bibitem{Laine:2021ego}
M.~Laine and S.~Procacci,
``Minimal warm inflation with complete medium response,''
JCAP \textbf{06} (2021), 031
[arXiv:2102.09913 [hep-ph]].
\bibitem{Visinelli:2011jy}
L.~Visinelli,
``Natural Warm Inflation,''
JCAP \textbf{09} (2011), 013
[arXiv:1107.3523 [astro-ph.CO]].
\bibitem{Kamali:2019ppi}
V.~Kamali,
``Warm pseudoscalar inflation,''
Phys. Rev. D \textbf{100} (2019) no.4, 043520
[arXiv:1901.01897 [gr-qc]].

\bibitem{Berera:1995ie}
A.~Berera,
``Warm inflation,''
Phys. Rev. Lett. \textbf{75} (1995), 3218-3221
[arXiv:astro-ph/9509049 [astro-ph]].
\bibitem{Berera:1995wh}
A.~Berera and L.~Z.~Fang,
``Thermally induced density perturbations in the inflation era,''
Phys. Rev. Lett. \textbf{74} (1995), 1912-1915
[arXiv:astro-ph/9501024 [astro-ph]].

\bibitem{Weinberg:1978kz}
S.~Weinberg,
``Phenomenological Lagrangians,''
Physica A \textbf{96} (1979) no.1-2, 327-340.
\bibitem{Gasser:1983yg}
J.~Gasser and H.~Leutwyler,
``Chiral Perturbation Theory to One Loop,''
Annals Phys. \textbf{158} (1984), 142.

\bibitem{Coleman:1969sm}
S.~R.~Coleman, J.~Wess and B.~Zumino,
``Structure of phenomenological Lagrangians. 1.,''
Phys. Rev. \textbf{177} (1969), 2239-2247.
\bibitem{Callan:1969sn}
C.~G.~Callan, Jr., S.~R.~Coleman, J.~Wess and B.~Zumino,
``Structure of phenomenological Lagrangians. 2.,''
Phys. Rev. \textbf{177} (1969), 2247-2250.
\bibitem{Bando:1987br}
M.~Bando, T.~Kugo and K.~Yamawaki,
``Nonlinear Realization and Hidden Local Symmetries,''
Phys. Rept. \textbf{164} (1988), 217-314.


\bibitem{Yokoyama:1998ju}
J.~Yokoyama and A.~D.~Linde,
``Is warm inflation possible?,''
Phys. Rev. D \textbf{60} (1999), 083509
[arXiv:hep-ph/9809409 [hep-ph]].


\bibitem{Dymnikova:2000gnk}
I.~Dymnikova and M.~Khlopov,
``Decay of cosmological constant as Bose condensate evaporation,''
Mod. Phys. Lett. A \textbf{15} (2000), 2305-2314
[arXiv:astro-ph/0102094 [astro-ph]].
\bibitem{Dymnikova:2001jy}
I.~Dymnikova and M.~Khlopov,
``Decay of cosmological constant in selfconsistent inflation,''
Eur. Phys. J. C \textbf{20} (2001), 139-146.

\bibitem{Levy:2020zfo}
M.~Levy, J.~G.~Rosa and L.~B.~Ventura,
``Warm inflation, neutrinos and dark matter: a minimal extension of the Standard Model,''
JHEP \textbf{12} (2021), 176
[arXiv:2012.03988 [hep-ph]].


\bibitem{Hall:2003zp}
L.~M.~H.~Hall, I.~G.~Moss and A.~Berera,
``Scalar perturbation spectra from warm inflation,''
Phys. Rev. D \textbf{69} (2004), 083525
[arXiv:astro-ph/0305015 [astro-ph]].
\bibitem{Ramos:2013nsa}
R.~O.~Ramos and L.~A.~da Silva,
``Power spectrum for inflation models with quantum and thermal noises,''
JCAP \textbf{03} (2013), 032
[arXiv:1302.3544 [astro-ph.CO]].
\bibitem{Bastero-Gil:2014jsa}
M.~Bastero-Gil, A.~Berera, I.~G.~Moss and R.~O.~Ramos,
``Cosmological fluctuations of a random field and radiation fluid,''
JCAP \textbf{05} (2014), 004
[arXiv:1401.1149 [astro-ph.CO]].
\bibitem{Visinelli:2014qla}
L.~Visinelli,
``Cosmological perturbations for an inflaton field coupled to radiation,''
JCAP \textbf{01} (2015), 005
[arXiv:1410.1187 [astro-ph.CO]].

\bibitem{Kohri:2016wof}
K.~Kohri and H.~Matsui,
``Higgs vacuum metastability in primordial inflation, preheating, and reheating,''
Phys. Rev. D \textbf{94} (2016) no.10, 103509
[arXiv:1602.02100 [hep-ph]].

\bibitem{Planck:2018vyg}
N.~Aghanim \textit{et al.} [Planck],
``Planck 2018 results. VI. Cosmological parameters,''
Astron. Astrophys. \textbf{641} (2020), A6
[erratum: Astron. Astrophys. \textbf{652} (2021), C4]
[arXiv:1807.06209 [astro-ph.CO]].


\end{thebibliography}
\end{document}